# Nonlinear optical analogies in quantum electrodynamics

D. H. Delphenich[†]
Physics Department
Bethany College
Lindsborg, KS USA 67456

*Abstract: Some of the basic notions of nonlinear optics are summarized and then applied to the case of the Dirac vacuum, as described by the Heisenberg-Euler effective one-loop Lagrangian. The theoretical and experimental basis for the appearance of nonlinear optical phenomena, such as the Kerr effect, Cotton-Mouton effect, and four-wave mixing are discussed. Further effects due to more exotic assumptions on the structure of spacetime, such as gravitational curvature and the topology of the Casimir vacuum are also presented.*

**1. Introduction.** Our main reasons for pursuing the possibility that the methods and concepts that are well established in nonlinear optics might be of use to the study of quantum electrodynamics are primarily the fact that quantum electrodynamics describes possible nonlinear interactions of photons with each other and with external electric or magnetic fields and the fact that the origin of these quantum corrections to linear electrodynamics seems to be in the polarizability of the electromagnetic vacuum. Now, by the term "electromagnetic vacuum," what we really intend is not a region of space in there is no energy present, whether in the form of mass or photons, but a region of space in which *only* an electromagnetic field is present [1]. Hence, there is some justification for treating the electromagnetic vacuum as a polarizable medium in the optical sense, which suggests that treating the electric permittivity and the magnetic permeability of the vacuum as simply constants, $\varepsilon_0$ and $\mu_0$, is basically a pre-quantum approximation, as well as the constancy of the speed of propagation of electromagnetic waves, $c_0 = 1/\sqrt{\varepsilon_0 \mu_0}$, or, equivalently, the index of refraction of the vacuum [2]. We shall regard these constants as asymptotic zero-field values of field dependent functions. The fact that $c_0$ itself might vary with the strength of the field suggests that quantum electrodynamics might even have something deep and subtle to say about causality itself that goes beyond the familiar concepts of special relativity.

From the standpoint of optics in general, it is then necessary to establish the electromagnetic properties of the vacuum when electromagnetic fields are present; at the very least, its constitutive law. Here, one must respect the limits of the phenomenological nature of the methods of quantum electrodynamics. Except for its treatment of bound states of charged particles, the vast majority of topics in quantum electrodynamics are treated in the scattering approximation. That is, the interactions of photons and charged particles are treated as particle scattering events from the outset, so the ultimate conclusion of the analysis is not really the solution of a system of field equations, as in classical electrodynamics, but the scattering amplitude – i.e., transition probability – for the scattering of the incoming particle states into outgoing particle states. Although these scattering amplitudes can take on considerable numerical accuracy from the consideration of successively higher-order (in powers of *e* or in numbers of loops) processes, such as radiative corrections, nonetheless, one must realize that the scattering process involves approximating the time interval during which the interaction took place as being much smaller that the time interval during which the scattering event took place (ideally, this interval is the entire time line). Hence, obtaining a more detailed picture of what happened in the actual interaction time interval is closely analogous to the way that geophysics tries to construct models of the Earth's interior by examining the way that seismic waves propagate from one point of the Earth's surface to another.

---

[†] delphenichd@bethanylb.edu

[1] Of course, if one considers the 2.7 K microwave background radiation that is present everywhere, this would always seem to be the case, but we shall consider such cosmological effects separately in the present discussion.

[2] From now on, we agree that the word "vacuum" will simply be a shortened form of "electromagnetic vacuum," which can include gravitational effects on the light cones.

However, one of the most powerful insights that one obtains from the successive approximations of the scattering amplitudes is due to the construction of "effective" models of vacuum structure. Generally, these models take the form of effective potentials and effective Lagrangians for the same processes, which represent corrections to the classical potentials and Lagrangians that account for quantum effects. For instance, it is common to construct effective one-loop potentials or Lagrangians, which is sometimes referred to as the *semi-classical* approximation.

The justification that we make for considering the nonlinear optical analogies for quantum electrodynamical processes is the possibility that what quantum electrodynamics is really showing us is how the linearity of Maxwell's equations breaks down in the realm of large field strengths. This is exactly the realm that one must consider when dealing with elementary charge distributions – e.g., nuclei and electrons – and the interactions of photons with such fields or with other photons of sufficiently high energy.

Our approach to defining the analogies between nonlinear optics and quantum electrodynamics will be to first summarize the relevant concepts that will be discussed in nonlinear optics in the first section, then briefly introduce the Dirac vacuum in the second section, and examine the extent to which one can regard it as an optical medium. We shall then discuss the Heisenberg-Euler Lagrangian, which represents a one-loop effective Lagrangian for the electromagnetic field when one includes vacuum polarization due to the formation of virtual electron-positron pairs, and the resulting nonlinear constitutive law that it defines. We shall then summarize the three main quantum electrodynamical processes that have suggested nonlinear optical analogies, namely, the interaction of photons with external electric fields, external magnetic fields, and other photons. We shall also mention the possible effect of background gravitational fields on the optical properties of the vacuum. Finally, we shall contrast the Casimir vacuum, which includes the zero-point electromagnetic field that one must consider in quantum electrodynamics, with the Dirac vacuum and discuss the resulting interaction of a photon with that zero-point field, which is referred to as the *Scharnhorst effect*.

**2. Nonlinear optical constitutive laws [1-4].** One of the formal differences between the methods of nonlinear optics and the methods of quantum electrodynamics that must be addressed is the fact that since optics in general is usually concerned with material media in the rest frame of the measuring devices, it is generally most straightforward to regard the electric and magnetic fields that collectively comprise an electromagnetic wave separately as time-varying vector fields $\mathbf{E}(t, \mathbf{x})$, $\mathbf{B}(t, \mathbf{x})$, in a three-dimensional space, while quantum electrodynamics, which attempts to maintain Lorentz invariance throughout, generally treats the electromagnetic field as a 2-form:

$$F = dt \wedge E + *B = E_i \, dt \wedge dx^i + \tfrac{1}{2} B^i \varepsilon_{ijk} \, dx^j \wedge dx^k \tag{2.1}$$

on Minkowski space. Hence, it is necessary to introduce a time + space – or "1 + 3" – decomposition L⊕Σ of Minkowski space into a one-dimensional time line L that represents the direction of proper time evolution for the rest space of the measuring devices and a complementary three-dimensional space Σ that describes that rest space.

Furthermore, since quantum electrodynamics is a gauge theory for the gauge group $U(1)$ it treats the chosen potential 1-form $A = A_\mu \, dx^\mu$ as the fundamental field, not $F = dA$. This is also necessary for the variational formulation of electromagnetism, since the field equations that one would obtain by varying only the 2-form $F$ in the electromagnetic field Lagrangian:

$$\mathcal{L}_{em} = \tfrac{1}{2} F \wedge *F = \frac{1}{4} F_{\alpha\beta} F^{\alpha\beta} dx^0 \wedge dx^1 \wedge dx^2 \wedge dx^3 \tag{2.2}$$

would be only $F = 0$, whereas varying $A$ produces the source-free Maxwell equation:

$$0 = \delta F = \partial_\mu F^{\mu\nu}. \tag{2.3}$$

The other Maxwell equation for $F$, viz., $dF = 0$, follows from the definition of $F$ as an exact form and the fact that $d^2 = 0$, in any event. In the gauge theory of electromagnetism, which treats the 1-form $A$ as the connection 1-form of a connection on a $U(1)$ principal bundle – viz., the gauge structure of the theory – and the 2-form $F = dA$ as its curvature 2-form, the latter identity becomes the Bianchi identity for the curvature.

The previous form of the source-free Maxwell equations assumes that the electromagnetic constitutive properties of the medium are simply those of the vacuum, and are therefore described by two constants – namely, $\varepsilon_0$ and $\mu_0$ – that are commonly set equal to unity to eliminate them from the qualitative consideration of the field equations. In optics, however, the constitutive laws become the center of consideration and cannot be neglected.

In general, if we denote the vector space of 2-forms on Minkowski space by $\Lambda^2$ then an electromagnetic constitutive law is a diffeomorphism $\kappa: \Lambda^2 \to \Lambda^2$, $F \mapsto G = \kappa(F)$; i.e., an invertible differentiable map that takes electromagnetic field strength 2-forms to electromagnetic excitation 2-forms and also has a differentiable inverse.

Actually, to be faithful to the concepts of optics, one must realize that since material media have finite time intervals during which they respond to applied electromagnetic waves and the response at one point of the medium will generally be influenced by the state of the medium at neighboring points of the medium, a linear electromagnetic constitutive law must be represented by a linear operator that takes vector fields to other vector fields and whose kernel is the impulse response matrix of the medium. Under Fourier transformation, the frequency-wave vector analog of this integral operator will take the form of a linear isomorphism from the complexification of $\Lambda^2$ to itself. If one ignores the absorptive properties of the medium, which are represented by the imaginary parts of the constitutive matrix in the frequency domain, the real part of the isomorphism is what we are calling $\kappa$, which therefore depends upon the frequency $\omega$ and wave number **k** of the field in it.

Hence, when one chooses a 6-frame $\{b^I, I = 1, \ldots, 6\}$ that spans $\Lambda^2$, so the 2-forms $F$ and $G$ take the form:

$$F = E_i\, b^i + B_i\, b^{i+3}, \qquad (2.4a)$$
$$G = D_i\, b^i + H_i\, b^{i+3}, \qquad (2.4a)$$

the constitutive map may be described by its matrix relative to this choice of framing:

$$[\kappa]^I_J = \left[\begin{array}{c|c} \varepsilon^i_j & \gamma^j_i \\ \hline \hat{\gamma}^i_j & (\mu^{-1})^i_j \end{array}\right], \qquad (2.5)$$

which gives the *E-B* form of the constitutive law [3]:

$$D_i = \varepsilon_{ij}\, E_j + \gamma_{ij}\, B_j, \qquad (2.6a)$$
$$H_i = \hat{\gamma}_{ij}\, E_j + (\mu^{-1})_{ij}\, B_j. \qquad (2.6b)$$

The $\varepsilon^i_j = \varepsilon^i_j(\omega, \mathbf{k})$ submatrix is referred to as the *electric permittivity matrix* of the medium, $(\mu^{-1})^i_j = (\mu^{-1})^i_j(\omega, \mathbf{k})$ is the inverse of the *magnetic permeability matrix*, and the matrices $\gamma^i_j = \gamma^i_j(\omega, \mathbf{k})$ and $\hat{\gamma}^i_j = \hat{\gamma}^i_j(\omega, \mathbf{k})$ are referred to as the *magneto-electric coupling matrices*. In most optical media, the latter two matrices are non-zero only due to the relative motion of the medium and the measuring devices, as in the case of the Fresnel-Fizeau effect, in which case, the matrices are real, or natural optical activity in the medium, in the complex case.

For nonlinear electromagnetic constitutive laws, it is more traditional to represent the diffeomorphism $\kappa$ by its "displacement vector field [4]":

$$Q = \frac{1}{4\pi}(G - F) = F = P_i\, b^i + M_i\, b^{i+3}, \qquad (2.7)$$

in which:

---

[3] From now on, we ignore the variance of the indices, due to the Euclidian character of the three-dimensional subspace of Minkowski space that we are dealing with, and write all indices as subscripts, as is conventional in classical optics.

[4] That is, it is a vector field on the vector space $\Lambda^2$, not on Minkowski space.

$$P_i = \frac{1}{4\pi}(D_i - E_i), \qquad M_i = \frac{1}{4\pi}(H_i - B_i), \tag{2.8}$$

are the components of the (electric) *polarization* and *magnetization* covectors, respectively. Although they take the form of the differences of fields, they are usually interpreted as the macroscopic densities of electric and magnetic dipole moments, respectively. This comes from the microscopic sense of the polarization of the medium then, namely, the formation of electric and magnetic dipoles in the medium under the action of an applied electric or magnetic field, respectively.

This makes the Maxwell equations take the form:

$$dF = 0, \qquad dF = -4\pi\, \delta Q, \tag{2.9}$$

and their **E**-**B** form becomes:

$$\nabla \cdot \mathbf{E} = -4\pi \nabla \cdot \mathbf{P}, \qquad \nabla \times \mathbf{E} + \frac{1}{c_0}\frac{\partial \mathbf{B}}{\partial t} = 0, \tag{2.10a}$$

$$\nabla \cdot \mathbf{B} = 0, \qquad \nabla \times \mathbf{B} - \frac{1}{c_0}\frac{\partial \mathbf{E}}{\partial t} = 4\pi\left(\frac{1}{c_0}\frac{\partial \mathbf{P}}{\partial t} + \nabla \times \mathbf{M}\right). \tag{2.10a}$$

The wave equation for *F* takes the Lorentz-invariant form:

$$\Box F = -4\pi\, d\delta Q, \tag{2.11}$$

in which $\Box = \delta d + d\delta$. This makes the wave equation for **E** become:

$$\Box \mathbf{E} = 4\pi\left[\nabla(\nabla \cdot \mathbf{P}) - \frac{1}{c_0}\frac{\partial}{\partial t}\left(\frac{1}{c_0}\frac{\partial \mathbf{P}}{\partial t} + \nabla \times \mathbf{M}\right)\right], \tag{2.12}$$

in which $\Box = 1/c_0^2\, \partial^2/\partial t^2 - \nabla^2$.

For optics, which is usually concerned with the magnetic field only insofar as it affects the dielectric properties of the medium, it is also traditional to expand the polarization covector field into a power series in *E*:

$$P_i(\omega, \mathbf{k}, \mathbf{E}) = \chi_i^{(0)}(\omega,\mathbf{k}) + \chi_{ij}^{(1)}(\omega,\mathbf{k})E_j + \chi_{ijk}^{(2)}(\omega,\mathbf{k})E_j E_k + \cdots \tag{2.13}$$

The coefficient arrays $\chi_{ij\ldots k}^{(n)}$, $n = 0, 1, \ldots$ are the $n^{\text{th}}$-order *electric susceptibilities*, which, like the successive derivatives in the Taylor series for *P*, are completely symmetric in their lower indices. It should be pointed out that this expansion is assumed to be valid in what nonlinear optics calls the regime of *weak nonlinearity*, which exists in the interval of field strengths between the breakdown of linearity in the response of the medium and the onset of phase transitions, such as electronic level transitions, in the medium that render the expansion invalid.

It is essential to observe that when a medium has a spatial inversion center this forces the polarization covector field to have only non-zero odd powers of the field **E**. In particular, the even-order susceptibilities must vanish.

The constant covector $\chi_i^{(0)}$ is the residual – or *static* – polarization of the medium, which usually vanishes, except for ferroelectric media. The first order susceptibility combines with the identity matrix to give the linear part of the constitutive law:

$$\varepsilon_{ij} = \varepsilon_0 (\delta_{ij} + \chi_{ij}^{(1)}). \tag{2.14}$$

The remaining sum of the terms for $n > 1$ gives the nonlinear part of the polarization covector field.

One generally approaches the nonlinear optical phenomena that emerge in the realm of large field strengths, such as high-intensity laser beams, by considering the effects at each order. In addition to looking at powers of a single applied field, one often considers the higher-order terms as giving nonlinear corrections to the linear superposition; i.e., wave mixing. The way that one achieves this in the context of the power series expansion (2.13) is to let:

$$\mathbf{E}(\omega, \mathbf{k}) = \mathbf{E}^{(1)}(\omega_1, \mathbf{k}_1) + \mathbf{E}^{(2)}(\omega_2, \mathbf{k}_2) + \ldots \tag{2.15}$$

and then regard each order $k$ of the series as being decomposable into a sum of terms of the form $\chi_{i \cdots j}^{(k)} E_i^{(l_1)} \cdots E_j^{(l_k)}$ which all have the same susceptibility tensor, although now it depends upon all of the associated frequencies and wave vectors.

Sometimes, when dealing with two fields, it is often useful to treat one of the waves as a background field while the other is an incoming wave, especially when one field is static. This generally results in making the susceptibilities functions of the background field. For instance, the possible coupling of a background magnetic field $\mathbf{B}_0$ to the susceptibilities is expressed by making them functions of $\mathbf{B}_0$, as well as $\omega$ and **k**:

$$P_i(\omega, \mathbf{k}, \mathbf{E}, \mathbf{B}_0) = \chi_i^{(0)}(\omega, \mathbf{k}, \mathbf{B}_0) + \chi_{ij}^{(1)}(\omega, \mathbf{k}, \mathbf{B}_0) E_j + \chi_{ijk}^{(2)}(\omega, \mathbf{k}, \mathbf{B}_0) E_j E_k + \cdots \tag{2.16}$$

In linear optics, one would expect, from superposition, that the result of combining two electromagnetic waves of frequencies $\omega_1$ and $\omega_1$ would be a polarization with only the original frequencies present. Indeed, in the absence of static polarization this is obvious from (2.13).

When one includes second order effects, which involve the products of two field strengths, one also finds contributions to the polarization that are produced at the sum and difference frequencies. In the event that both incoming waves have the same frequency, this gives a d.c. field, as well as the double frequency of the incoming waves. The former effect is referred to as *optical rectification* and the latter is called *second harmonic generation*.

Whenever one has a situation in which the total number of incoming and outgoing waves is greater than two, one must add a condition that is the wave-mechanical analogue of the conservation of energy-momentum – up to a factor of $\hbar$ – namely, that if $\omega_i$, $\mathbf{k}_i$, $i = 1, \ldots N$ are the frequencies and wave vectors of the waves, with a + or − sign according to whether they are incoming or outgoing waves, resp., then:

$$\sum_i \omega_i = 0, \qquad \sum_i \mathbf{k}_i = 0. \tag{2.17}$$

One refers to this condition as the *phase matching* condition. In the case of second harmonic generation, since the frequencies satisfy $2\omega_1 = \omega_2$, phase matching demands that the wave vectors must satisfy $2\mathbf{k}_1 = \mathbf{k}_2$.

If one treats one of two waves – say, $\mathbf{E}_0$ – as a background field then this makes the second order term linear in the other – say $\mathbf{E}$. One can then combine the second order term with the linear term to give, to first order:

$$\mathbf{P} = (\varepsilon + \chi^{(2)}(\mathbf{E}_0))\mathbf{E} = \varepsilon(\mathbf{E}_0)\mathbf{E}, \tag{2.18}$$

which says that the dielectric tensor – hence, the principal indices of refraction − is being modulated by the background field $\mathbf{E}_0$, which is known as the *linear electro-optic*, or *Pockels effect*.

In third order, one has a nonlinear analogue of the Pockels effect, in which one modulates $\varepsilon$ with the third-order susceptibility when evaluated on $\mathbf{E}_0 \otimes \mathbf{E}_0$:

$$\mathbf{P} = (\varepsilon + \chi^{(3)}(\mathbf{E}_0, \mathbf{E}_0))\mathbf{E} = \varepsilon(\mathbf{E}_0)\mathbf{E}, \tag{2.19}$$

which is known as the *quadratic electro-optic*, or *Kerr effect*. One also encounters *four-wave mixing*, in which one can also produce waves of frequency $\omega_3 = 2\omega_1 - \omega_2$. Hence, phase matching demands that one must have $\mathbf{k}_3 = 2\mathbf{k}_1 - \mathbf{k}_2$.

Many of the most optically important nonlinear effects concern media in which there are resonant frequencies, such as electronic transition frequencies and phonon modes. We shall only briefly mention some of the possible resonance effects. With *Raman scattering*, the incoming wave of frequency $\omega$ mixes with a vibrational mode $\omega_v$ of the medium to produce waves of frequencies $\omega - \omega_v$ and $\omega + \omega_v$, which are referred to as the *Stokes* and *anti-Stokes* components of the scattered wave, respectively; the Stokes component is what produces the visible light when a fluorescent material is exposed to ultraviolet light. Similarly, *Brillouin scattering* results when the incoming wave mixes with acoustic modes that produce an effective diffraction grating in the medium.

One can also encounter solitonic effects, such as self-transparency of a medium when the incoming wave is close to an electronic transition frequency, and self-focusing of a cylindrical beam due to the fact that the index of refraction varies with the intensity of the beam, which varies with the distance from the beam center. The former effect brings into play the sine-Gordon equation, while the latter one involves the nonlinear Schrödinger equation.

There are effects produced by a background magnetic field that are analogous to those produced by a background electric field. When the polarization takes the form (2.16), one has corresponding linear and quadratic magneto-optic effects, corresponding to the effect of $\mathbf{B}_0$ on the second and third order electric susceptibilities, respectively. In fact, one must distinguish between the modulation of the index of refraction by $\mathbf{B}_0$ for incoming waves whose direction of propagation $\mathbf{k}$ is parallel or perpendicular to the direction of $\mathbf{B}$. It is the former case, the result is a Faraday rotation of the polarization 2-frame (i.e., the orthonormal frame $\{\mathbf{\epsilon}_1, \mathbf{\epsilon}_2\}$ that is composed of the unit vectors in the directions of oscillation of the $\mathbf{E}$ and $\mathbf{B}$ vectors).  One also has the conversion of circularly polarized waves to elliptically polarized ones upon reflection, which is referred to as the *magneto-optic Kerr effect* [5].  In the latter case, the presence of non-zero magnetization makes an otherwise isotropic medium behave like a uniaxial dielectric, so there will be different indices of refraction depending on whether $\mathbf{E}$ is parallel or perpendicular to $\mathbf{B}_0$, which is called the *Cotton-Mouton effect*.  Hence, the presence of $\mathbf{B}_0$ can make a medium *birefringent* at sufficiently high field strengths for a wave propagating in it. (A medium is birefringent when the index of refraction depends upon the direction of polarization of a wave that is propagating in it, in the sense of the angular orientation of the $\{\mathbf{\epsilon}_1, \mathbf{\epsilon}_2\}$ frame in the plane of polarization.)

### 3. The Dirac vacuum [5-7].

The key physical process at the root of most quantum electrodynamics is the creation of a particle/anti-particle pair from a photon and its inverse process of pair annihilation. It becomes immediately clear that there will be a minimum photon energy for pair creation to be possible that corresponds to the rest energy of the least massive particles that can be produced, namely, an electron and a positron.  This critical energy is then $2m_e c^2$, which is roughly 1 MeV, and which corresponds to a photon whose frequency is $2m_e c^2/h = 3 \times 10^{20}$ Hz, which puts the photon in the low gamma part of the spectrum. The next possible pair that could be created is the muon/anti-muon pair, which takes place at 211 MeV, and requires $6 \times 10^{22}$ Hz photons, which would be further into the gamma range.  Between those critical energies, the difference between the photon energy and the critical energy for electron/positron creation would show up as the total kinetic energy of the particles produced.  Beyond the critical energy for muon/anti-muon production, one quickly approaches the 270 MeV threshold for the creation of a neutral pion and its anti-particle.  However, since pions are strongly interacting, this also defines the limits of applicability for the methods of quantum electrodynamics.

From conservation of momentum, pair creation is not a spontaneous decay process for a sufficiently energetic photon, but something that only takes place in the presence of an external field, such as a nuclear electrostatic field.  However, the same is not true for pair annihilation, which can happen as a result of a simple collision.

Below the critical energy for electron/positron production, one can still think of the photon as producing a *virtual* electron/positron pair, which can be perhaps thought of as equating the photon with a bound state of an electron and positron.  One can think of a region of space in which there are many such photons as exhibiting *vacuum polarization*. Hence, this is what we will use as the fundamental basis for the analogy between nonlinear optics and some quantum electrodynamical processes, namely, the

---

[5] Note the possible source of confusion in the term "Kerr effect," which hopefully should be resolved by the context of its usage.

consideration of a polarized vacuum as a sort of nonlinear optical medium, which we call the *Dirac vacuum.*

In addition to the polarization of photon fields at high energy, one also expects the vacuum to become polarized in regions of very high static electric field strengths, such as in the small neighborhoods of elementary charge distributions, such as electrons and nuclei. This suggests that the most fundamental issue is not so much that of a critical field energy, but that of a critical field strength. The value that is most commonly used is the field strength at the "surface of the classical electron," i.e., the Coulomb value at a distance from a point charge *e* that equals the classical electron radius. This value is:

$$E_c = \frac{m_e^2 c^3}{e\hbar}, \tag{3.1}$$

which equals either $1.3 \times 10^{16}$ V/cm or $4.4 \times 10^{13}$ G. Presumably, at the surface of the classical electron, vacuum polarization sets in and invalidates the classical model. Hence, one can think of a Dirac vacuum as existing either in the near vicinity of an elementary charge or in a region of space occupied by a high-energy photon.

For the sake of analogy, the regime of weak nonlinearity in the Dirac vacuum then exists in the interval of field strengths between the breakdown of linearity due to vacuum polarization and the point at which the primary phase transition of pair creation begins, which is defined by the critical field strength.

Although the formation of electric dipoles due to a presumably infinitesimal spatial separation of the particle pair in its bound state seems clear enough, as we shall see in the next section, the bound state is also associated with a non-zero magnetic dipole moment, which apparently originates in the fact that both particles in the pair have spin ½ since the photon has spin one.

## 4. Heisenberg-Euler constitutive law.

In 1936, Heisenberg and Euler [**8**] derived what is now described as a one-loop effective Lagrangian for the electromagnetic field by starting with Dirac's theory of positrons. Their calculations were confirmed by Weisskopf [**9**] and Schwinger [**10**] using other methods of derivation. This Lagrangian then defined the starting point for more detailed studies of the electrodynamics of the Dirac vacuum that followed, some of which we will discuss in the next section. (For a recent survey of Heisenberg-Euler Lagrangians and their modifications, see Dunne [**11**].)

The Heisenberg-Euler Lagrangian, being assumed Lorentz invariant and gauge invariant, is best expressed in terms of the only Lorentz and gauge invariant functions of the electromagnetic field strength 2-form *F* that exist, namely:

$$\mathscr{F} = \tfrac{1}{2} F_{\mu\nu} F^{\mu\nu} = E^2 - B^2, \qquad \mathscr{G} = \tfrac{1}{2} F_{\mu\nu} {}^*F^{\mu\nu} = -2\mathbf{E} \cdot \mathbf{B}. \tag{4.1}$$

One notes that this makes the Lagrangian for the Maxwellian electromagnetic field *in vacuo* equal to:

$$\mathcal{L}_{em} = \tfrac{1}{2} \mathcal{F} \, . \tag{4.2}$$

In this expression, we are rescaling the **E** and **B** vectors to make $\varepsilon_0 = \mu_0 = 1$.

The Heisenberg-Euler Lagrangian is the sum of this classical – also called "zero-loop" or "tree level" – Lagrangian and an effective one-loop correction term:

$$\mathcal{L}_1 = \frac{\alpha}{2\pi} \int_0^\infty d\eta \, \frac{e^{-\eta}}{\eta^3} \left\{ (E_c^2 - \frac{\eta^2}{3} \mathcal{F}) - i\eta^2 \mathcal{G} \, \frac{\cos\left(\frac{\eta}{E_c}\sqrt{\mathcal{F} - i\mathcal{G}}\right) + c.c.}{\cos\left(\frac{\eta}{E_c}\sqrt{\mathcal{F} - i\mathcal{G}}\right) - c.c.} \right\}, \tag{4.3}$$

in which $\alpha = e^2/\hbar c \approx 1/137$ is the fine-structure constant.

The weak-field approximation in this case is defined by field strengths $E$ or $B$ that are much less than the critical field strength. One can then expand (4.3) in a power series, which we give to sixth order in the field strengths:

$$\mathcal{L}_1 \approx \xi \left\{ \frac{1}{2}\left(\mathcal{F}^2 + 7\mathcal{G}^2\right) + \frac{1}{7E_c^2}(13\mathcal{G}^2\mathcal{F} + 2\mathcal{F}^3) \right\}, \tag{4.4}$$

in which we have set $\xi = \alpha/180\pi E_c^2 \approx 10^{-42}$ cm$^2$/V$^2$, which gives one some idea of the relative order of magnitude of the nonlinear correction compared to the conventional Lagrangian.

There is also a strong-field approximation that applies to fields much stronger than the critical value, but since the present discussion is only concerned with optical phenomena, we shall not need the explicit form of its expression. It is however, illuminating to write out (4.4) in its *E-B* form explicitly:

$$\mathcal{L}_1 \approx \xi \left\{ \frac{1}{2}\left[(E^2 - B^2)^2 + 7(\mathbf{E}\cdot\mathbf{B})^2\right] \right. \\ \left. + \frac{1}{7E_c^2}[13(\mathbf{E}\cdot\mathbf{B})^2(E^2 - B^2) + 2(E^2 - B^2)^3] \right\}. \tag{4.5}$$

In order to obtain the Lorentz-invariant form of the electromagnetic constitutive law that follows from this Lagrangian, one can use the fact that (cf., [**12**]):

$$G_{\mu\nu} = \frac{\partial \mathcal{L}}{\partial F^{\mu\nu}} = \frac{\partial \mathcal{L}}{\partial \mathcal{F}}\frac{\partial \mathcal{F}}{\partial F^{\mu\nu}} + \frac{\partial \mathcal{L}}{\partial \mathcal{G}}\frac{\partial \mathcal{G}}{\partial F^{\mu\nu}} = \frac{\partial \mathcal{L}}{\partial \mathcal{F}} *F_{\mu\nu} + \frac{\partial \mathcal{L}}{\partial \mathcal{G}} F_{\mu\nu} \, . \tag{4.6}$$

For the Heisenberg-Euler Lagrangian $\mathcal{L} = \mathcal{L}_{em} + \mathcal{L}_1$, with the one-loop correction expressed in the form (4.4), this gives:

$$G_{\mu\nu} = \xi \left\{ \left[ \mathcal{F} + \frac{1}{7E_c^2}(13\mathcal{G}^2 + 6\mathcal{F}^2) \right] * F_{\mu\nu} + \mathcal{G} \left[ 7 + \frac{26}{7E_c^2}\mathcal{F} \right] F_{\mu\nu} \right\}. \tag{4.7}$$

It is also useful to have the constitutive law in its *E-B* form, which one derives from:

$$D_i = \frac{\partial \mathcal{L}}{\partial E^i} = \frac{\partial \mathcal{L}}{\partial \mathcal{F}} \frac{\partial \mathcal{F}}{\partial E^i} + \frac{\partial \mathcal{L}}{\partial \mathcal{G}} \frac{\partial \mathcal{G}}{\partial E^i} = 2 \frac{\partial \mathcal{L}}{\partial \mathcal{F}} E_i - 2 \frac{\partial \mathcal{L}}{\partial \mathcal{G}} B_i, \tag{4.8a}$$

$$H_i = -\frac{\partial \mathcal{L}}{\partial B^i} = -\frac{\partial \mathcal{L}}{\partial \mathcal{F}} \frac{\partial \mathcal{F}}{\partial B^i} - \frac{\partial \mathcal{L}}{\partial \mathcal{G}} \frac{\partial \mathcal{G}}{\partial B^i} = 2 \frac{\partial \mathcal{L}}{\partial \mathcal{F}} B_i + 2 \frac{\partial \mathcal{L}}{\partial \mathcal{G}} E_i. \tag{4.8b}$$

If we compare this to (2.6a, b) this gives our basic matrices as:

$$\varepsilon_{ij} = \frac{\partial \mathcal{L}}{\partial \mathcal{F}} \delta_{ij} =$$

$$= \left\{ 1 + 4\xi \left[ 2(E^2 - B^2) + \frac{1}{7E_c^2}(13(\mathbf{E}\cdot\mathbf{B})^2 + 6(E^2 - B^2)^2) \right] \right\} \delta_{ij}, \tag{4.9a}$$

$$\gamma_{ij} = -\frac{\partial \mathcal{L}}{\partial \mathcal{G}} \delta_{ij} = 4\xi (\mathbf{E}\cdot\mathbf{B}) \left[ 7 + \frac{26}{7E_c^2}(E^2 - B^2) \right] \delta_{ij}, \tag{4.9b}$$

$$\hat{\gamma}_{ij} = -\gamma_{ij}, \tag{4.9c}$$

$$(\mu^{-1})_{ij} = 2 \frac{\partial \mathcal{L}}{\partial \mathcal{F}} \delta_{ij} = \varepsilon_{ij}. \tag{4.9d}$$

In this form, we can regard our Dirac vacuum as a *bi-isotropic* medium since all of the submatrices above are of the form $f(\mathbf{E}, \mathbf{B})\delta_{ij}$. In particular, one should note that the dependency of $\mathcal{L}_1$ on $\mathcal{G}$ has generated non-zero magneto-electric coupling terms. Furthermore, the symmetric and anti-symmetric parts of the 6×6 constitutive matrix $\kappa(\mathbf{E}, \mathbf{B})$ are of the form:

$$^{(1)}\kappa = \varepsilon(\mathbf{E},\mathbf{B}) \begin{bmatrix} \delta_{ij} & 0 \\ 0 & \delta_{ij} \end{bmatrix}, \qquad ^{(2)}\kappa = \gamma(\mathbf{E},\mathbf{B}) \begin{bmatrix} 0 & \delta_{ij} \\ -\delta_{ij} & 0 \end{bmatrix}, \tag{4.10}$$

with the obvious definitions for the functions $\varepsilon(\mathbf{E}, \mathbf{B})$ and $\gamma(\mathbf{E}, \mathbf{B})$. In these expressions, we are using the notation of Hehl and Obukhov [4], who refer to the matrices $^{(1)}\kappa$ and $^{(2)}\kappa$ as the *principal part* and the *skewon part* of the matrix $\kappa$, respectively. The *axion* part, which is proportional to the volume element, vanishes.

In order to identify the nonlinear electric susceptibilities, we first write down the components of the polarization covector, to third order in **E**, as:

$$P_i(\mathbf{E}, \mathbf{B}) = \frac{\xi}{\pi} \left[ -2B^2 E_i + 7 B_i B_j E_j + 2 \delta_{jk} E_i E_j E_k \right]. \tag{4.11}$$

By symmetrizing the last coefficient, which we denote by parentheses, we can identify the nonlinear electric susceptibilities as:

$$\chi_i^{(0)} = 0 \,, \tag{4.12a}$$

$$\chi_{ij}^{(1)} = -\frac{\xi}{\pi}[2B^2 \delta_{ij} - 7 B_i B_j]\,, \tag{4.12b}$$

$$\chi_{ijk}^{(2)} = 0, \tag{4.12c}$$

$$\chi_{ijkl}^{(3)} = \frac{2\xi}{\pi} \delta_{i(j} \delta_{kl)}\,, \tag{4.12d}$$

Although optics usually ignores the magnetic susceptibilities, in the case of the Dirac vacuum there is a duality symmetry ($\mathbf{E} \rightarrow \mathbf{B}$, $\mathbf{B} \rightarrow -\mathbf{E}$) in the way that the $\mathbf{E}$ and $\mathbf{B}$ fields appear that is useful in analyzing the effects of background fields. Hence, we write the magnetization vector to third order in the form:

$$M_i(\mathbf{E},\mathbf{B}) = \frac{\xi}{\pi}\Big[ 2E^2 B_i - 7 E_i E_j E_j - 2 \delta_{jk} B_i B_j B_k \Big]. \tag{4.13}$$

This gives a corresponding set of magnetic susceptibilities:

$$\zeta_i^{(0)} = 0 \,, \tag{4.14a}$$

$$\zeta_{ij}^{(1)} = \frac{\xi}{\pi}[2E^2 \delta_{ij} - 7 E_i E_j]\,, \tag{4.14b}$$

$$\zeta_{ijk}^{(2)} = 0, \tag{4.14c}$$

$$\zeta_{ijkl}^{(3)} = \frac{2\xi}{\pi} \delta_{i(j} \delta_{kl)}\,, \tag{4.14d}$$

Some general features of the Dirac vacuum can then be read off by inspection:
  *i*) The residual polarization and magnetization both vanish, as one might expect.
  *ii*) The roles of $\mathbf{E}$ and $\mathbf{B}$ are interchangeable (up to sign) in the susceptibilities, which follows from the duality invariance of the Lagrangian.
  *iii*) The polarization (magnetization, resp.) is an odd functions of $\mathbf{E}$ ($\mathbf{B}$, resp.), as one might also expect; i.e., reversing $\mathbf{E}$ ($\mathbf{B}$, resp.) reverses the resulting polarization (magnetization, resp.).
  *iv*) Due to the fact that the third-order electric susceptibility is independent of $\mathbf{B}$, even a static $\mathbf{E}$ field by itself will generate a non-vanishing polarization of the vacuum for field strengths close enough to $E_c$ ; an analogous remark is true for the magnetization. This essentially comes from the one-loop correction to Coulomb's law.
  *v*) From (4.11), the polarization and magnetization both vanish for a plane electromagnetic wave, which has the property that $\mathcal{F} = \mathcal{G} = 0$.

When one goes to field strengths that exceed the critical value, one can no longer use the weak-field Lagrangian, and must use a different way of expression it. The net effect on the optical properties is to introduce non-zero imaginary contributions to the electric permittivity and magnetic permeability, along with the index of refraction. This has the effect of introducing the absorption of some of the incoming wave energy due to the creation of electron/positron pairs.

### 5. Photons propagating in the Dirac vacuum [13-34].

Now that we have the nonlinear electric susceptibilities, we examine them for the nonlinear optical phenomena that each might imply at each successive order. We can consider three basic states of the Dirac vacuum: the presence of an external **B** field, the presence of an external **E** field, and the presence of an electromagnetic wave. For each basic state, we will then examine how that state affects the propagation of an electromagnetic wave in the medium. In the last case, we are then dealing with the nonlinear superposition of two high-intensity photons, or photon-photon scattering.

From the order of magnitude of the critical field strength, one sees that it is not likely that such electric or magnetic fields are possible in any conventional laboratory. Hence, one must look to atomic sources of high electric field strengths, such as the small neighborhoods of nuclei and electrons, and astronomical source of high magnetic field strengths, such as the vicinity of pulsars and magnetars, whose magnetic fields can be within an order of magnitude of the critical value. Furthermore, we shall assume that the wavelength of the photons considered is much less than the Compton wavelength, which gives an order of magnitude for their spatial extent, and any external fields are slowly-varying in space and time.

*a. Photon propagating in an external magnetic field.* From the form of the nonlinear electric susceptibilities, it is clear that this state is the most direct to consider, and we regard the **B** field that appears in them as the external magnetic field. Furthermore, without loss of generality, we can regard the **B** field as pointing in the *z*-direction, so, if we disregard the self-interaction of the incoming electromagnetic wave, the electric permittivity takes the form:

$$\varepsilon(B^2) = \begin{bmatrix} \varepsilon_\perp(B^2) & 0 & 0 \\ 0 & \varepsilon_\perp(B^2) & 0 \\ 0 & 0 & \varepsilon_\parallel(B^2) \end{bmatrix}, \tag{5.1}$$

in which:

$$\varepsilon_\perp(B^2) = 1 - \frac{2\alpha}{45\pi}\left(\frac{B}{B_c}\right)^2, \tag{5.2a}$$

$$\varepsilon_\parallel(B^2) = 1 + \frac{\alpha}{9\pi}\left(\frac{B}{B_c}\right)^2. \tag{5.2b}$$

When one applies Fresnel analysis (cf., Brezin and Itzikson [**13**]) to this with:

$$\mu_{ij}^{-1}(B^2) = \left(1 - \frac{2\alpha B^2}{45\pi E_c^2}\right)\delta_{ij}. \tag{5.3}$$

one arrives at a Fresnel quartic hypersurface that decomposes into two quadratic surfaces in space, whose solution vectors **n** represent the possible directions of propagation (so **n** is collinear with **k**) for an outgoing plane wave, and whose lengths $n$ equal the index of refraction in that direction. If $\theta$ is the angle between **k** and **B** and **ε** is the polarization vector of the wave then we can describe the two propagation modes as:

$$\text{transverse:} \quad n_\perp(\mathbf{B}) = \sqrt{1 + \frac{4\alpha}{45\pi}\left(\frac{B}{B_c}\right)\sin^2\theta}, \quad \boldsymbol{\varepsilon}_\perp = \mathbf{n} \times \mathbf{B}, \tag{5.4a}$$

$$\text{parallel:} \quad n_\|(\mathbf{B}) = \sqrt{1 + \frac{7\alpha}{45\pi}\left(\frac{B}{B_c}\right)\sin^2\theta}, \quad \boldsymbol{\varepsilon}_\| = \mathbf{B} - (\mathbf{B}\cdot\mathbf{n})\times\mathbf{B}, \tag{5.4b}$$

in which the terms "transverse" and "parallel" refer to the orientation of **E** with respect to the plane spanned by {**k**, **B**).

Hence, since the presence of **B** has given the permittivity tensor (5.1) the form of a uniaxial dielectric, and thus produced birefringence, we see that quantum electrodynamics – to one loop order – predicts a Cotton-Mouton effect. The wave that propagates in the transverse mode is the *ordinary* wave and the one that propagates in the parallel mode is the *extraordinary* wave.

Furthermore, if the incoming wave is linearly polarized with an initial polarization:

$$\boldsymbol{\varepsilon}(t) = (\alpha_1\,\boldsymbol{\varepsilon}_1 + \alpha_2\,\boldsymbol{\varepsilon}_2)\cos\omega t \tag{5.5}$$

and **k** perpendicular to **B** then after it has traveled a distance $L$ it will be elliptically polarized with:

$$\boldsymbol{\varepsilon}(t) = \alpha_1\,\boldsymbol{\varepsilon}_1\cos(\omega t - n_1 L) + \alpha_2\,\boldsymbol{\varepsilon}_2\cos(\omega t - n_2 L) \tag{5.6}$$

which is the Kerr effect, with an associated Faraday rotation.

In order to get some numerical sense of the effect, one can compute for the *Kerr constant:*

$$\frac{n_\perp - n_\|}{\lambda B^2} \approx \frac{7\alpha}{90\pi\lambda E_c^2}, \tag{5.7}$$

in which $\lambda$ is the wavelength of the incoming wave. For the 500 nm line of sodium vapor, this gives a value of the Kerr constant of about $2\times 10^{-34}$ m/V$^2$, which is twenty orders of magnitude smaller that the Kerr constant for water $5\times 10^{-14}$ m/V$^2$. This is consistent with the fact that the nonlinear quantum electrodynamical effects are actually quite difficult to experimentally observe.

When the **B** field is strong enough to cause pair formation from the oncoming photon, we can no longer use the weak-field Lagrangian to obtain the permittivities and permeabilities, but must work with the Heisenberg-Euler Lagrangian directly. As the analytical details are quite involved, but well documented (cf., [**14-22**] [6]), we shall only summarize the results qualitatively. First, the $\varepsilon_{ij}$ and $\mu_{ij}$ tensors become complex, which also implies that the transverse and parallel indices of refraction $n_\perp$ and $n_\parallel$ will be complex, as well. In addition to the difference between the real parts of $n_\perp$ and $n_\parallel$ that gives the birefringence of the vacuum in the presence of a field, there will also be a difference between the imaginary parts, which is called *dichroism*. Since the imaginary parts represent the absorption of energy by the formation of particle pairs, this amounts to the statement that when the incoming photon is polarized transverse to the **B** field, it will decay into a particle pair, which amounts to an effective absorption of the photon by the medium, whereas if it is polarized parallel to the field then there will be no decay; hence, no absorption. Thus, a magnetized vacuum will act just as a Polaroid filter does by absorbing photons that are not polarized along the preferred direction. However, the actual numerical magnitude of the effect is comparable to that of the aforementioned birefringence.

In the range of field strengths that are close to the critical value, but not greater than it, yet another optical process becomes possible: an incoming photon can split into two slower photons. The treatment of this process in the literature is usually quantum electrodynamical [**5-7**], in that one first obtains the relativistic scattering amplitude for the process from Feynman diagram techniques, and from that derives the transition rate, absorption length, and absorption coefficients $\kappa_\perp$ and $\kappa_\parallel$ (viz., the inverse absorption length). This gives the imaginary part of the indices of refraction and the real parts $\kappa_\perp$ and $\kappa_\parallel$ can then be obtained from the Kramers-Kronig relations.

As it was announced in Adler, et al [**18**], and calculated in detail in [**19**], the only scattering process that is not forbidden kinematically for small values of the parameter:

$$\beta = \frac{\omega}{m_e} \frac{B \sin \theta}{B_c}. \tag{5.8}$$

or by CP invariance is the one in which a photon that is polarized parallel [7] to the **k-B** plane decays into two nearly-parallel photons that are polarized transverse to it. This tends to suppress "box" diagrams, so the leading contribution to the scattering amplitude comes from "hexagon" diagrams. We denote these two scattering channels by the subscripts 1 and 2, respectively. The absorption coefficients are obtained by [**14**] are, in units of 1/cm:

$$\kappa_1 = 0.12 \left(\frac{\omega}{m_e}\right)^5 \left(\frac{B \sin \theta}{B_c}\right)^6, \tag{5.9a}$$

---

[6] Note that the statement in Klein and Nigam [**16**] that strong magnetic fields do not produce particle pairs from photons is no longer regarded as true (cf., Erber [**22**]).

[7] N.B. In Adler, et al, [**17**] it is the **B** vector of the photon that defines its polarization.

$$\kappa_2 = 0.39 \left(\frac{\omega}{m_e}\right)^5 \left(\frac{B \sin\theta}{B_c}\right)^6, \tag{5.9b}$$

and the indices of refraction for parallel and transverse-polarized incoming photons are, for photons of sub-critical energy:

$$n_{\parallel} = 1 + \frac{\alpha}{\pi}\left(\frac{B \sin\theta}{B_c}\right)^2 (0.18 + 0.24 \xi^2), \tag{5.10a}$$

$$n_{\perp} = 1 + \frac{\alpha}{\pi}\left(\frac{B \sin\theta}{B_c}\right)^2 (0.31 + 0.44 \xi^2). \tag{5.10b}$$

Hence, medium is still birefringent as a result of the presence of the magnetic field, as it was for the simple propagation of a single photon.

The possibility of experimentally testing the phenomenon of magnetic birefringence of the vacuum is being developed by the PVLAS facility in Italy [**23**]. They plan to use a 110 mW laser that produces photons with a wavelength of 1.062 $\mu$m and sends them through a vacuum between the poles of an 8 T superconducting dipole magnet. The effect that will ultimately be measured is the ellipticity of the polarization that results from the birefringence.

### b. Photon propagating in an external electric field.

Due to the duality symmetry of the susceptibilities, the effects of an external **E** field are analogous to those of an external **B** field, except that the modes have reversed, since the **E** field modulates the magnetic permeabilities, not the electric permittivity. That is, this time one has principal magnetic permeabilities of
:

$$\mu_{\parallel}^{-1}(E^2) = 1 - \frac{\alpha}{9\pi}\left(\frac{E}{E_c}\right)^2, \qquad \mu_{\perp}^{-1}(E^2) = 1 + \frac{2\alpha}{45\pi}\left(\frac{E}{E_c}\right)^2 \tag{5.11}$$

an electric permittivity of:

$$\varepsilon_{ij}(E^2) = 1 + \frac{2\alpha}{45\pi}\left(\frac{E}{E_c}\right)^2 \delta_{ij}. \tag{5.12}$$

and the resulting modes are:

$$\text{transverse:} \quad n_{\perp}(\mathbf{E}) = \sqrt{1 + \frac{7\alpha}{45\pi}\left(\frac{E}{E_c}\right)^2 \sin^2\theta}, \qquad \boldsymbol{\varepsilon}_{\perp} = \mathbf{n} \times \mathbf{E}, \tag{5.13a}$$

$$\text{parallel:} \quad n_{\parallel}(\mathbf{E}) = \sqrt{1 + \frac{4\alpha}{45\pi}\left(\frac{E}{E_c}\right)\sin^2\theta}, \qquad \boldsymbol{\varepsilon}_{\parallel} = \mathbf{E} - (\mathbf{E}\cdot\mathbf{n})\times\mathbf{E}. \tag{5.13b}$$

Hence, one also has electro-optic Kerr and Cotton-Mouton effects due to the presence of an external **E** field, as well as dichroism and photon splitting.

The effects of vacuum polarization on the propagation of photons in a strong electric field were the first to be verified experimentally, since such effects were predicted by Delbrück [**24**] in the scattering of gamma rays from heavy nuclei. Both the birefringence and photon splitting were verified experimentally by Jarlskog, et al [**25**].

*c. Photon-photon scattering*. In linear Maxwellian electrodynamics the superposition of field solutions to produce field solutions prohibits any interaction between photons following intersecting rays. Hence, one of the earliest consequences of the nonlinearity in the Heisenberg-Euler electromagnetic wave equations to be pointed out was the possibility that at high enough field strengths the formation of virtual electron/positron pairs might imply an actual scattering of photons by other photons.

In the case of virtual pair formation, which involves total field strengths that are sub-critical, one refers to the process as *elastic* photon-photon scattering. When actual pairs are formed by super-critical total field strengths, however, since one does not count the kinetic energy of the actual pairs in the energy of the scattered photons (here, we are envisioning a beam that consists of a large number of photons), one then refers to such a process as *inelastic* photon-photon scattering.

The quantum electrodynamical computation of the transition amplitude for elastic photon-photon scattering was first carried out by Euler [**26**] and then refined by others, such as Karplus and Neuman [**27**]. However, it was immediately observed that the resulting total scattering cross-section was actually quite negligible for photons in the optical frequency range, about $0.7 \times 10^{-65}$ cm$^2$ (cf., e.g., [**6**]).

Although this fact tended to discourage early experimentalists, nevertheless, the fact that vacuum polarization effects, such as birefringence and photon splitting, had been experimentally observed in the context of Delbrück scattering kept alive the belief that eventually photon-photon scattering might be observed in the laboratory.

In more recent decades, advances in laser technology and astrophysics re-kindled the fire of enthusiasm amongst experimenters that photon-photon scattering might actually be observable. The technique that we shall discuss in this study is the one that pertains to nonlinear optics the most closely, namely, the technique of four-wave mixing with second harmonic generation. However, we shall still mention that some of the other techniques that have been proposed involve resonance enhancement of the interaction in microwave waveguide cavities [**28**] and plasmas [**29**]. Of particular interest in the latter case are the pair plasmas, such as plasmas composed of electron/positron pairs, which are expected to be found in the magnetospheres of neutron stars and magnetars.

One notes that since the invariants $\mathscr{F}$ and $\mathscr{G}$ vanish for a plane wave – or even a superposition of plane waves with parallel propagation vectors – the only way that the Heisenberg-Euler correction to the classical electromagnetic Lagrangian can be non-vanishing is if the field is non-planar, such as the superposition of two or more non-parallel plane waves.

However, the scattered signal produced by two incoming plane waves is generally governed by the first-order susceptibilities and is still too weak to be measured experimentally. Hence, it was suggested by Rozanov [**30**] and developed by others [**31-34**], that perhaps one could enhance the scattered wave by means of a third field. In Ding and Kaplan [**31**], the third field was a static magnetic field, but later it took the form of a third plane wave. This then produces the four-wave mixing that was mentioned above. The phase matching condition, with three incoming plane waves with wave vectors $k_i$, $i = 1, 2, 3$, and frequencies $\omega_i$, $i = 1, 2, 3$, and an outgoing plane wave of wave vector $k_4$ and frequency $\omega_4$ is then essentially the wave analog of energy-momentum conservation at a vertex that one would expect of the analogous Feynman diagram:

$$\mathbf{k}_1 + \mathbf{k}_2 = \mathbf{k}_3 + \mathbf{k}_4 \qquad (5.14a)$$
$$\omega_1 + \omega_2 = \omega_3 + \omega_4. \qquad (5.14b)$$

Due to the non-vanishing cubic susceptibilities of the Dirac vacuum, one gets second-harmonic generation in the outgoing wave. This has the experimental advantage that the frequency of the scattered wave is sufficiently distinct from those of the incoming waves that its detection is simpler than when one must separate closely-spaced frequencies.

So far, a previous attempt [**35**] at experimentally observing photon-photon scattering by means of four-wave mixing was unsuccessful. However, a project that is currently underway [**36**] involving the Astrid Gemini system at the Rutherford Appleton Laboratory at Oxford expects that such a demonstration might be possible in 2007.

*d. Self-focusing.* Another fundamental point of departure between linear Maxwellian electrodynamics and quantum electrodynamics is found in the elementary notion that the photon is generally assumed to be a stable spatially and temporally localized manifestation of electromagnetic wave energy. This seems hard to justify in linear electrodynamics because any attempt localize the support of a wave – such as modulating a plane wave with an envelope function of compact spacetime support – will necessarily imply a spectrum of frequencies and wave numbers, and in linear electrodynamics any dispersion in a medium will cause spreading of the wave packet in space and time. Hence, it seems clear that a linear theory of the vacuum cannot account for the stability of localized electromagnetic wave packets.

As observed above, in a nonlinear medium, it is possible for there to also be self-focusing of the wave packet, which tends to contradict the tendency of the packet to spread. When the relevant parameters are such that the two processes cancel precisely, one obtains a *soliton* solution of the nonlinear wave equation.

In the papers of Rozanov [**37**], and later Soljačić and Segev [**38**], it was shown that theoretically such a situation was possible in the case of the Dirac vacuum. However, since the Heisenberg-Euler nonlinear corrections vanish for a plane wave, or even a plane wave modulated by a slowly-varying envelope, what they proposed was to use a pair of plane waves intersecting at right angles. They showed that for such a configuration if the combined wave was expressible in the form:

$$\mathbf{E}(t, x, y, z) = \frac{1}{2} A(x)[\cos(\mathbf{k}\cdot\mathbf{r} - \omega t)\hat{\mathbf{n}} + \cos(\underline{\mathbf{k}}\cdot\mathbf{r} - \omega t)\underline{\hat{\mathbf{n}}}], \qquad (5.15)$$

in which $\hat{\mathbf{n}}$ is any unit vector and the underbar signifies the mirror image of a vector in the *xt*-plane, then the use of the Heisenberg-Euler corrected wave equation for **E**, with suppression of third-order harmonics, implies that the envelope function $A(x)$ must satisfy the cubic nonlinear Schrödinger equation:

$$\frac{d^2 A}{dx^2} + \Gamma A + \frac{\xi}{2k^2} A^3 = 0, \qquad (5.16)$$

in which:

$$\Gamma = \left(\frac{\omega}{c_0}\right)^2 - k_y^2 - k_z^2 \qquad (5.17)$$

is the nonlinear dispersion relation and $\xi$ is the coupling constant in the Heisenberg-Euler Lagrangian.

Since this nonlinear Schrödinger equation, which is widely used in nonlinear optics, is known to admit solitonic solutions, the authors of [**37, 38**] propose that such a configuration of crossed laser beams might possibly produce stable solitons. However, they point out that although the beam intensity that would be necessary – viz., $10^{31}$ W/cm$^2$ – is beyond the state of the art in laser technology, they anticipate that it should be reachable in the foreseeable future.

*e. Gravitational effects.* So far, we have assumed that the spacetime manifold is Minkowski space. One naturally wonders what the effect might be if one goes to a more general Lorentz manifold. In particular, since the Lorentzian metric is responsible for defining the light cones in each tangent space that restrict the way that elect6romagnetic waves might propagate, one would expect that a curved-space metric would have to have some effect on the propagation of photons, beyond the deflection of null rays that is discussed in general relativity.

In the paper of Drummond and Hathrell [**39**], this possibility is explored, with the result that it is the curvature that couples to the electromagnetic 2-form $F$, and in such a way that when the curvature is anisotropic, one can deduce a sort of "gravitational birefringence" that, moreover, can also imply that the value of $c_0$ can be greater than its vacuum value for the extraordinary wave.

Rather than the Heisenberg-Euler contribution to the standard electromagnetic Lagrangian, they start with a general expression that involves the Riemann curvature tensor $R_{\mu\nu\rho\sigma}$, the Ricci tensor $R_{\mu\nu}$, and the scalar curvature $R$ coupled to $F_{\mu\nu}$:

$$\mathscr{L}_{\mathrm{DH}} = \frac{1}{m_e^2}(aR\mathscr{F} + bR_{\mu\nu}F^{\mu\sigma}F^\nu{}_\sigma + cR_{\mu\nu\sigma\tau}F^{\mu\nu}F^{\sigma\tau} + d\,D_\mu F^{\mu\nu} D_\alpha F^\alpha{}_\nu), \qquad (5.18)$$

in which $D_\mu$ refers to the covariant differential using the Levi-Civita connection.

Only the last term is present in flat space and the authors show that the value for $d$ that one derives from the flat-space vacuum polarization amplitude is of order $e^2$ and may be neglected in the rest of the analysis. The values of $a$, $b$, and $c$ are obtained by coupling a graviton to two on-mass-shell photons in flat space:

$$(a, b, c) = -\frac{\alpha}{720\pi}(5, -26, 2). \tag{5.19}$$

One can then think of these scalars as the gravitational form factors of the photon.

When one derives the field equations from the electromagnetic Lagrangian with the above correction, the Bianchi identity is unchanged, but the Maxwell equation becomes:

$$D_\mu F^{\mu\nu} + \frac{1}{m_e^2}[4aR\,F^{\mu\nu} + 2b(R^\mu{}_\sigma F^{\sigma\nu} - R^\nu{}_\sigma F^{\sigma\mu}) + 4c\,R^{\mu\nu}{}_{\sigma\tau} F^{\sigma\tau}] = 0. \tag{5.20}$$

Since the curvature is coupled to the electromagnetic field in this equation, Drummond and Hathrell then examine the effect of this coupling for the various cosmological models that are used in general relativity. For instance, in cases where the curvature is constant, such as the de Sitter spacetime, equation (5.20) reduces to simply a scalar multiple of the conventional Maxwell equation, and one would expect no observable change in the propagation of photons.

In a gravitational wave background, one obtains two values of $c$, depending upon whether the photon is propagating parallel or anti-parallel to the gravitational wave. If we call then $c_+$ and $c_-$, resp., then one has:

$$c_+ = c_0, \qquad c_- = (1 + 2b\xi^2\omega^2)\,c_0, \tag{5.21}$$

in which $\xi$ is the Heisenberg-Euler coupling constant and $\omega$ is the frequency of the gravitational wave. Although the correction to $c_0$ in $c_-$ is immeasurably small, it is nonetheless positive, which suggests many deep questions to be resolved regarding quantum causality in this sense.

In the Schwarzschild background, $c$ will be different for photons propagating in the radial and transverse directions relative to the gravitating center.

$$c_r = (1 + \tfrac{1}{2}\varepsilon)\,c_0, \qquad c_t = (1 - \tfrac{1}{2}\varepsilon)\,c_0, \tag{5.22}$$

in which $\varepsilon = 6MG\xi^2/r^3$.

For the Robertson-Walker spacetime, the spatial isotropy and homogeneity combine to make the value of $c$ slightly smaller, but still isotropic. However, when one goes to the Friedman modification – i.e., filling the universe with a homogeneous isotropic fluid – the change in $c$ is positive and increases as $1/t^2$ as one approaches the Big-Bang singularity.

### 6. The Casimir vacuum and the Scharnhorst effect.

The Casimir vacuum [**40, 41**] differs from the Dirac vacuum only in that the background electromagnetic field that one imposes is the "zero-point" field that quantum electrodynamics predicts must exist as a result of quantum fluctuations of the electromagnetic vacuum state. This is essentially an infinite-dimensional extension of the fact that the quantum harmonic oscillator has a non-zero ground state $\hbar\omega_n$ due to the uncertainty principle if one regards the quantum electromagnetic field as a continuous distribution of coupled quantum harmonic oscillators. The basic mathematical principle that brings about the Casimir effect is the fact that the zero-point field must vanish on the conductors that define the boundary of a region in which the zero-point energy is non-vanishing, even though the total field energy between them is non-vanishing.

The existence of a zero-point field between two parallel uncharged perfect conducting plates was theoretically predicted by Casimir [**40, 41**] and experimentally verified by Lamoreaux [**42**], as well as Mohideen and Anushree [**43**]. If the plates are separated by a distance $L$ in the $x$ direction then the zero-point energy density (per unit area of a plate) is given by:

$$U_{zp}(\mathbf{r}) = -\frac{\pi^2}{720 L^3}. \tag{6.1}$$

The zero-point field that corresponds to this is then:

$$\mathbf{E}_{zp}(\mathbf{r}) = -\nabla U_{zp} = -\frac{\pi^2}{240 L^4}\mathbf{i}, \tag{6.2}$$

whose units are those of pressure.

The 2-loop effective Lagrangian for the zero-point field takes the form:

$$\mathcal{L}_{zp} = \tfrac{1}{2}(\varepsilon_{ij}E^i E^j - \mu_{ij}^{-1}B^i B^j), \tag{6.3}$$

in which both tensors are diagonal and:

$$\varepsilon_{ij}(L, x^3) = \delta_{ij} + \frac{\pi^2}{180}\frac{\alpha^2}{(m_e L)^4}\begin{bmatrix} f(x)-\frac{11}{45} & 0 & 0 \\ 0 & f(x)-\frac{11}{45} & 0 \\ 0 & 0 & f(x)+\frac{11}{45} \end{bmatrix}, \tag{6.3a}$$

$$\mu_{ij}(L, x^3) = \delta_{ij} - \frac{\pi^2}{180}\frac{\alpha^2}{(m_e L)^4}\begin{bmatrix} f(x)+\frac{11}{45} & 0 & 0 \\ 0 & f(x)+\frac{11}{45} & 0 \\ 0 & 0 & f(x)-\frac{11}{45} \end{bmatrix}. \tag{6.3b}$$

The function $f(x)$ that appears in the matrices will not be specified further in this discussion since it does not appear in the indices of refraction in the one-loop approximation. Once again, we are looking at the permittivities and permeabilities of a uniaxial dielectric.

In the low-frequency limit, an incoming photon will see indices of refraction parallel and transverse to the plates that are given by:

$$n_{\parallel}(0) = 1, \qquad n_{\perp}(0) = 1 - \frac{11\pi^2}{8100} \frac{\alpha^2}{(m_e L)^4} \,. \tag{6.4}$$

This birefringence of the Casimir vacuum, which is referred to as the *Scharnhorst effect* [**39, 40**], is analogous to the birefringence that one would expect from a background electric field, except that it produces a reduction of the index of refraction and it is only applicable to sub-critical incoming photons. However, considering the comments made above about the low order of magnitude for the effect as a result of a near-critical electric field, one can reasonably imagine the effect is even more immeasurable for a field as weak as the zero-point field.

Since a reduction in the index of refraction to a value below one suggests that the propagation of a transverse photon will be superluminal, one might wonder what happens in the opposite limit of high-frequency, which corresponds to the geometrical optics approximation that is closer to the formalism of relativity theory. One can use the Kramers-Kronig relation to show that:

$$n(\infty) = n(0) - \frac{2}{\pi} \int_0^\infty \frac{d\omega'}{\omega'} \operatorname{Im} n(\omega') \,. \tag{6.5}$$

Although an explicit form for *n* as a function of $\omega$ is still forthcoming, nevertheless, one can assume that it is a positive function, which means that $n_{\perp}(0) > n_{\perp}(\infty)$. Hence, in the geometrical optics limit the photons will be even more superluminal.

### Acknowledgements

The author wishes to thank Friedrich Hehl, Guillermo Rubilar, and Yakov Itin for illuminating discussions concerning various aspects of the topics treated in this work, and to thank Bethany College for providing a congenial and unhurried environment in which to do that work.